\documentclass[prl,twocolumn,showpacs,amsmath,amssymb,superscriptaddress]{revtex4}
\usepackage{graphicx}
\begin{document}
\title{Conductance Characteristics between a Normal Metal and a Superconductor\\
Carrying a Supercurrent}
\author{Degang Zhang}
\author{C. S. Ting}
\affiliation{Texas Center for Superconductivity and Department of Physics,\\
University of Houston, Houston, TX 77204, USA}
\author{C.-R. Hu}
\affiliation{Department of Physics, Texas A\& M University, 
College Station, Texas, 77843, USA}
%
%
%\date{\today}
\begin{abstract}
The low-temperature conductance (G) characteristics between a normal 
metal and a clean superconductor (S) carrying a supercurrent 
$I_s$ parallel to the interface is theoretically investigated.
Increasing $I_s$ causes lowering and broadening of (1) coherence 
peaks of s-wave S, and d-wave S at (100) contact, 
(2) midgap-states-induced zero-bias conductance peak for d-wave
S at (110) contact, and (3) Andreev-reflection-induced enhancement
of $G$ within the gap near the metallic-contact limit. Novel features 
found include a current-induced central peak and a three-humped 
structure at intermediate barrier strength, etc.  

\end{abstract}
\maketitle
It is well-known that Andreev reflection plays a fundamental
role in understanding the transport properties of a normal
metal/superconductor junction (NSJ) [1]. From the
current-voltage ($I$-$V$), or the differential conductance 
[$G(V)\equiv dI(V)/dV$] characteristics of the junction, 
one can learn much information about S, includng its 
elementary excitation spectrum and its order-parameter 
symmetry, etc. Blonder et al. have developed a general theory [2]
for studying $I$-$V$ and $G(V)$ of an NSJ that allows a 
dimensionless barrier-strength parameter $z$ to range from 
metallic contact, $z=0$, to the tunneling regime, $z>>1$. 
There only conventional $s$-wave symmetry for S was considered. 
Recently, much attention has been paid to the conductance
characteristics of d-wave, cuprate S in both theory and 
experiment [3-13]. Due solely to the sign change of the 
d-wave gap-function order parameter $\Delta({\bf k})$ on the Fermi 
surface, a zero-bias conductance peak (ZBCP) appears in the 
tunneling spectrum of an N/(d-wave S) junction with 
non-(n0m) contact [4,5,6]. The ZBCP arises from a sizable number 
of midgap states formed at the S side of the N/S interface and 
appears for all $z$ but is narrower and taller for larger $z$. 
In a large magnetic field, the ZBCP splits into 
two peaks [7,8,10,11]. It is interesting to also study the 
effect of a supercurrent $I_s$ in S on $G(V)$. Very 
recently, $G(V)$ for tunneling into a diffusive s-wave 
superconducting  wire carrying an $I_s$ was measured and compared 
with theory [14]. It was shown that the coherence peaks were 
suppressed and broadened with increasing $I_s$, and the 
effect is the same as that caused by a magnetic field. 
Because the width and thickness of the wire were smaller than 
the superconducting coherence length and penetration depth, all 
variations transverse to the wire could be neglected, as was the 
magnetic field generated by $I_s$. 
The positions of the coherence peaks in $G(V)$ were found to 
practically not shift with $I_s$, up to $\sim 4/5$ 
of the critical current. In this work, we investigate theoretically
the conductance characteristics of a clean NSJ with an $I_s$ in S 
parallel to the interface by extending the theory of 
Blonder et al. [2]. Contrary to Ref. [14], this work is not limited 
to large $z$. We consider both $s$-wave and $d$-wave S with (100) 
and (110) contacts. Some novel results are obtained, especially for 
$z\stackrel{<}{\sim}1$, when $G(V)$ does not simply reflect 
the thermally-smeared quasi-particle density of states. Hopefully, 
these predictions can be confirmed experimentally. Unlike Ref. 
[14], the present work does not consider Coulomb blockade, which 
is presumably not so important in an extended system and in the clean 
limit. As in Ref. [14], we also assume a uniform $I_s$, 
and neglect self-field.

When a uniform $I_s$ passes through a conventional
three-dimensional s-wave S, the phase of $\Delta({\bf k})$ 
has a spatial variation of $2{\bf q}_s\cdot {\bf x}$, 
where ${\bf x}$ is the center-of-mass 
position of a Cooper pair, ${\bf q}_s = (m_*/2){\bf v}_s$, 
with ${\bf v}_s$ the supercurrent velocity, and $m_*$ the 
mass of a Cooper pair. ($\hbar = 1$ is assumed throughout 
this work.) At temperature $T = 0$, the magnitude of the order 
parameter $\Delta_q$ stays unchanged until the Landau criterion is 
satisfied (i.e. $q = 0.5\Delta^0$, where $q\equiv q_s/k_F$ and 
$\Delta^0\equiv \Delta_0 /E_F$). Here $\Delta_0$ 
is the superconducting gap when $I_s = 0$, 
$k_F$ and $E_F$ are the Fermi momentum and energy, 
respectively. When $q\ge 0.5\Delta^0$, S
becomes gapless, and quasiparticles are generated in a 
portion of the Fermi surface [15]. We shall see that 
this can lead to a ZBCP in $G(V)$ for an N/(s-wave S)
junction with the barrier strength $z\simeq 1$. 
This current-induced ZBCP is always quite broad and not very 
tall, and its height decreases for larger $z$. It is therefore 
characteristically different from the ZBCP induced by the midgap 
surface states in a d-wave S with non-(n0m) contact that 
is narrower and taller for larger $z$ [4]. (The midgap-states-induced 
ZBCP has been ubiquitously observed in high-$T_c$ cuprate and 
other unconventional Ss.) 

As $q$ is increased further, $\Delta_q$  
gradually decreases to zero at $q=0.67\Delta^0$
[Fig. 1(a)]. The supercurrent density quickly reaches a peak 
(the thermodynamic critical current density) at $q = q_c = 
0.515\Delta^0$ [Figs. 1(c)] [16]. The region $q>q_c$, 
in which $I_s$ is a decreasing function of $q$, 
is unstable and can not be observed experimentally. 
(For a two-dimensional s-wave S, superconductivity disappears 
immediately after the Landau criterion is met. Then 
$q_c = 0.5\Delta^0$.)

\begin{figure}
\rotatebox[origin=c]{180}{\includegraphics[angle=90, 
           height=1.6in]{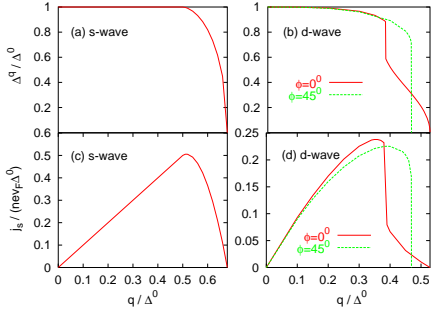}}
\caption {\label{fig:orderparam} Dependence of the superconducting 
order parameter on the normalized supercurrent-velocity parameter 
$q$ for (a) an s-wave and (b) a d-wave S. ($\phi$ is 
the angle between the supercurrent and the antinodal direction in 
the latter case.) In (c) and (d), the corresponding dependences of 
supercurrent density on $q$ are given.}
\end{figure}

Different from that in an s-wave S, the $\Delta_q$-vs-$q$ 
relation in a d-wave S also depends on the direction of the 
supercurrent. (Here $\Delta_q$ denotes the maximum gap in the 
presence of $I_s$.) For a two-dimensional d-wave 
S with a supercurrent, the gap-function order 
parameter at $T = 0$ is described by [17]
$$ \pi {\rm ln}\frac{\Delta^0}{\Delta^q}=\int_{\ge} d\theta
{\rm cos}^2(2\theta) {\rm ln} (g+\sqrt{g^2-1}),\eqno{(1)}$$
where $g\equiv \frac{2q}{\Delta^q}|\frac{{\rm cos}(\theta-\phi)}
{\cos(2\theta)}|$, $\Delta^q=\Delta_q/E_F$, $\phi$ is the angle 
between the supercurrent and the antinodal direction, and the 
integral in Eq. (1) is from 0 to $2\pi$ with the constraint
$g^2-1\ge 0$. 

Figure 1(b) shows the dependence of the d-wave $\Delta_q$
on $q$ at $\phi = 0$ and ${\pi}/{4}$. We can see that when 
$q$ is less than $\sim 0.3\Delta^0$, the changes of the 
order parameter with $q$ in both the antinodal and nodal 
directions are almost the same. However, a great difference
exists  for larger $q$. When $I_s$ is applied along 
the antinodal direction, $\Delta^q$ has a sharp drop 
(from $0.883\Delta^0$ to $0.588\Delta^0$) between $q=0.384\Delta^0$
and $0.385\Delta^0$. After that it drops continuously 
to zero at $q=0.53\Delta^0$. When $\phi=\pi/4$,
$\Delta^q$ gradually decreases to $0.689\Delta^0$
at $q=0.469\Delta^0$, and has no solution beyond. Fig. 1(d) 
gives the corresponding dependences of the supercurrent 
density on $q$ [17]. It is seen that the thermodynamic critical 
current is reached at $q = q_c = 0. 35\Delta^0$ ($0.39\Delta^0$) 
for current in the antinodal (nodal) direction. 

The elementary excitations in S are governed
by the time-independent Bogoliubov-de Gennes equations [18]:
$$
E u({\bf x})=h_0 u({\bf x})+\int d{\bf x}^\prime\Delta
     ({\bf s},{\bf r})v({\bf x}^\prime)\,,\eqno{(2a)}$$
$$
E v({\bf x})=-h_0 v({\bf x})+\int d{\bf x}^\prime\Delta^*
     ({\bf s},{\bf r})u({\bf x}^\prime)\,, \eqno{(2b)}
      $$
where ${\bf s}={\bf x}-{\bf x}^\prime, {\bf r}=\frac{1}{2}
({\bf x} + {\bf x}^\prime)$, and $h_0=-\frac{{\bf \nabla}
^2}{2m} +U\delta(x) - \mu$ with $\mu$ the chemical potential. 
It is useful to express the superconducting order parameter 
in the form: $\Delta
({\bf s},{\bf r})=\int d{\bf k}e^{i{\bf k}\cdot{\bf s}}
\bar{\Delta}({\bf k},{\bf r})e^{i2{\bf q}_s\cdot{\bf r}}$ [3].
Neglecting the proximity effect near the N/S interface
at $x=0$, we have $\bar{\Delta}({\bf k},{\bf r})
= \Delta ({\bf k})\Theta(x)$, where $\Theta(x)$ is a 
step function, and $\Delta ({\bf k})$ is the order parameter
of a bulk S in the presence of $I_s$.

In the WKBJ approximation, Eqs. (2) have special solutions
of the form
$$
\left (
\begin{array}{c}
u\\
v
\end{array}\right ) = e^{i{\bf k}_F\cdot {\bf x}}
\left (
\begin{array}{c}
e^{i{\bf q}_s\cdot {\bf x}} \bar u\\
e^{-i{\bf q}_s\cdot {\bf x}} \bar v
\end{array}\right )\,,
\eqno{(3)}
$$
where $\bar u ({\bf x})$ and $\bar v ({\bf x})$ obey
the generalized Andreev equations [1]:
$$
(E-\frac{{\bf q}_s^2}{2m}-\frac{{\bf q}_s\cdot {\bf k}_F}{m})\bar u\\ 
=-\frac{i({\bf k}_F+{\bf q}_s)}{m}\cdot {\bf \nabla} \bar u
  +\Delta ({\bf k}_F)\Theta(x)\bar v\,, \eqno{(4a)
}
$$
$$
(E+\frac{{\bf q}_s^2}{2m}-\frac{{\bf q}_s\cdot {\bf k}_F}{m})\bar v\\
 = \frac{i({\bf k}_F-{\bf q}_s)}{m}\cdot {\bf \nabla} \bar v
   +\Delta^*({\bf k}_F)\Theta(x)\bar u\,. \eqno{(4b)
}
$$
Obviously, the eigenenergy $E$ is symmetric about 
$E = {\bf q}_s\cdot {\bf k}_F/m$. When ${\bf q}_s$ 
is applied parallel to the interface of the NSJ, i.e. 
${\bf q}_s=-q_s{\bf e}_y$, we have
$$\left (
\begin{array}{c}
\bar u_\nu\\
\bar v_\nu
\end{array}\right )=e^{i\alpha_\nu x}\left (
\begin{array}{c}
u_\nu^>\\
v_\nu^>
\end{array}\right )~~~ ({\rm for} ~ x>0)\,,\eqno{(5a)} $$
$$\left (
\begin{array}{c}
\bar u_\nu\\
\bar v_\nu
\end{array}\right )=\left (
\begin{array}{c}
e^{i\beta_\nu x}u_\nu^<\\
e^{i\gamma_\nu x}v_\nu^<
\end{array}\right )~~~ ({\rm for} ~ x<0)\,,\eqno{(5b)}$$
where $\nu= {\mathrm sign}(k_{Fx})$; 
$\alpha_\nu = [-\nu{\bf q}_s^2/2+mA_\nu]/|k_{Fx}|$, with 
$A_\nu\equiv\sqrt{(E+q_sk_{Fy}/m)^2-\Delta_\nu({\bf k}_F)
\Delta^*_\nu({\bf k}_F)}$;
$\beta_\nu = m\nu[-{\bf q}^2_s/(2m)
+ E+q_sk_{Fy}/m]/|k_{Fx}|$; $\gamma_\nu = -m\nu
[{\bf q}^2_s/(2m) + E + q_sk_{Fy}/m]/|k_{Fx}|$;
$u_\nu^{>(<)}$ and $v_\nu^{>(<)}$ are constants.
For example, in S, we have $B_\nu\equiv u_\nu^>/
v_\nu^> = \Delta_\nu({\bf k}_F)/(E+q_sk_{Fy}/m - \nu A_\nu)$.

Following Ref. [2], after a tedious but straightforward 
calculation, we obtain the Andreev and normal reflection 
coefficients, $a(E)$ and $b(E)$:

\begin{widetext}
$$a(E)=\frac{2q_+(k_++k_-)}{B_-(-k_-+q_++2imU)
(k_+-q_-+2imU)-B_+(k_++q_++2imU)(-k_--q_-+2imU)}\,,\eqno{(6a)}$$
$$b(E)=\frac{B_+(k_-+q_--2imU)
(-k_++q_+-2imU)+B_-(k_+-q_-+2imU)(k_-+q_+-2imU)}
{B_-(-k_-+q_++2imU)
(k_+-q_-+2imU)-B_+(k_++q_++2imU)(-k_--q_-+2imU)}\,.\eqno{(6b)}$$
\end{widetext}

Here $q_+=|k_{Fx}|+\beta_+$, $q_-=|k_{Fx}|+\gamma_+,$
and $k_\nu=|k_{Fx}|+\nu \alpha_\nu$. The critical
supercurrent velocity is much less than the Fermi velocity.
So the Andreev approximation, $q_\pm\approx k_\pm
\approx |k_{Fx}|$, also holds in the presence of a supercurrent. 
The normalized conductance can then be calculated according 
to a formula given in Ref. [2]:
\begin{widetext}
$$ G=\frac{G_s}{G_n}\,,\;\;\;\;\;\;
G_n=-\frac{e^2}{\pi}\int^{+\infty}_{-\infty}dE
\int^{\frac{\pi}{2}}
_{-\frac{\pi}{2}}d\theta \frac{\partial f(E-eV)}{\partial E}
[1-|b(+\infty)|^2]\,,$$
$$G_s=-\frac{e^2}{\pi}\int^{+\infty}_{-\infty}dE
\int^{\frac{\pi}{2}}
_{-\frac{\pi}{2}}d\theta \frac{\partial f(E-eV)}{\partial E}
[1+|a(-E)|^2-|b(E)|^2]\,,\eqno{(7)}$$
\end{widetext}
where $|k_{Fx}|=k_F {\rm cos}\theta$, $f(E)$ is the Fermi
distribution function, $G_n$ and $G_s$ are the differential 
conductance for S in the normal and superconducting states, respectively.

\underline{{\it S-wave superconductor}}. In this case, the superconducting
order parameter $\Delta_\nu ({\bf k}_F) =\Delta_q$ is independent 
of $\nu$.

\begin{figure}
\rotatebox[origin=c]{180}{\includegraphics[angle=90, 
           height=1.6in]{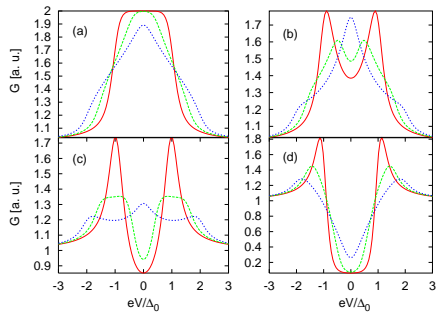}}
\caption {\label{fig:swave} The normalized differential conductance 
vs voltage for a normal metal/s-wave superconductor junction: 
(a) $z=0$, (b)$z=0.5$, (c) $z=1.0$, and (d) $z=5.0$. Red: $q=0$, 
green: $q=0.3\Delta^0$, and blue: $q=0.515\Delta^0$, at which the 
thermodynamic critical current is reached. Note that $q\equiv q_s/k_F$, 
and $\Delta^0\equiv\Delta_0/E_F$.}
\end{figure}

In Fig. 2, $G(V)$ at various $q$ and $z\equiv 2mU/k_F$ is plotted. 
[We have used $k_BT = 0.01 E_F$ and $\Delta_0 = 0.1E_F$.]
When $z=0$ and $q=0$, electrons incoming with all momenta ${\bf k}_F$ 
with $k_{Fx}>0$ can enter S and equal number of holes at opposite momenta 
are retro-reflected into N if
$|eV|<\Delta_0$. So the normalized conductance $G = 2.0$ within the
superconducting gap if $T=0$. With increasing $q$, the range of $G=2.0$
diminishes and the $G(V)$ curve turns into a nearly triangular peak 
centered at zero bias [Fig. 2(a)]. At large $z$ [Fig. 2(d)], the 
coherence peaks are suppressed and broadened with increasing $q$, 
but contrary to the case of a diffusive superconducting 
wire [14], here the peaks of $G(V)$ move outward while the gap shrinks.   
The intermediate-$z$ results are even richer in behavior [Figs.2(b) 
and (c)]: A fairly broad and not very tall peak appears at zero bias 
and a three-humped structure can also appear for nearly critical $q$. 
Note that the larger is $z$, the lower is this current-induced ZBCP. 
The area under this peak is also not conserved as $z$ changes. 
These features are characteristically different from the ZBCP 
induced by the midgap surface states in d-wave S with 
non-(n0m) contacts. [4]       

For electrons entering an NSJ at a fixed incident angle $\theta$,
a ZBCP would result from their contributions to the normalized 
conductance if $2q|\sin\theta|>\Delta^0$ is satisfied. Thus, one can 
see this peak only if $q > 0.5 \Delta^0$ is satisfied.
For $0.5\Delta^0 < q < 0.67\Delta^0$, there is a critical angle 
$|\theta _c|=\arcsin(\Delta^0/2q)$, which decreases from 90$^{\circ}$ 
to $48.3^{\circ}$ in this range. No ZBCP is induced by 
electrons with incident angle $|\theta|<|\theta_c|$. However, 
only a small portion of this regime can be observed, because 
only the region $q \le 0.515\Delta^0$ is stable. 

\underline{{\it D-wave superconductor}}. In this case, the pair 
potential has the form $\Delta_\nu ({\bf k}_F)=\Delta_q {\rm cos}
(2\theta_\nu)$. Here, $\theta_\nu=\theta+\nu \alpha$, $\alpha$ is 
the angle between the antinodal direction and the positive $x$ 
axis.

\begin{figure}
\rotatebox[origin=c]{180}{\includegraphics[angle=90, 
           height=1.6in]{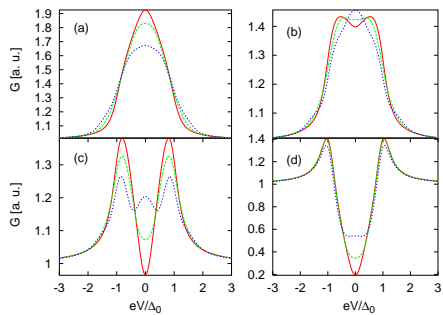}}
\caption {\label{fig:d100} The normalized differential conductance 
vs voltage for a normal metal/d-wave superconductor junction with 
(100) contact: (a) $z=0$, (b)$z=0.5$, (c) $z=1.0$, and (d) $z=5.0$.
Red: $q=0$, green: $q=0.2\Delta^0$, and blue: $q=0.35\Delta^0$, at 
which the thermodynamic critical current is reached.}
\end{figure}

Figure 3 presents the normalized conductance at different $z$ and $q$ 
for a d-wave S with (100) contact (i.e. $\alpha=0^0$). 
For $z=0$ [Fig. 3(a)], the central peak due to Andreev 
reflection is gradually suppressed and slightly broadened. For large 
$z$ [Fig. 3(d)] one sees mainly the filling up of the central dip with 
only a slight suppression of the coherence peaks as $q$ increases. 
For intermediate $z$ [Figs. 3(b) and (c)], one sees intricate 
behavior with some similarity to the corresponding figures in Fig. 2.

\begin{figure}
\rotatebox[origin=c]{180}{\includegraphics[angle=90, 
           height=1.6in]{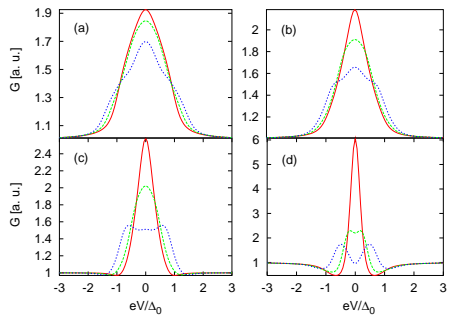}}
\caption {\label{fig:d110} The normalized differential conductance 
vs voltage for a normal metal/d-wave superconductor junction with 
(110) contact. The $z$ values considered are the same as in Fig. 3.
The $q$ values considered are: Red: $q=0$, green: $q=0.2\Delta^0$, 
and blue: $q=0.39\Delta^0$, at which the critical current is reached.}
\end{figure}

Figure 4 shows the normalized conductance at different $z$ and $q$ 
for a d-wave S with (110) contact (i.e. $\alpha=45^0$).
It is seen that the ZBCP induced by the midgap surface states 
is suppressed, broadened, and eventually split at sufficiently large 
$z$ when $q$ is increased. 

In conclusion, we have studied the differential conductance
of a clean normal metal/superconductor junction carrying a 
supercurrent parallel to the junction interface, for barrier 
strength ranging from metallic-contact to the tunneling regime. 
In the tunneling regime, we obtain results
similar to the case of a diffusive s-wave superconducting wire 
studied recently, viz., suppression and broadening of the 
coherence peaks for both an s-wave superconductor and a d-wave 
superconductor with (100) contact, except that the coherence 
peaks are found to move outward in the s-wave case.
For d-wave superconductor with (110) contact we also find 
the midgap-surface-states-induced ZBCP to be suppressed and 
broadened and eventually split with increasing supercurrent. 
In the metallic-contact limit, supercurrent causes the 
Andreev-reflection-induced conductance enhencement within
the (maximum) gap to become weakened and broadened.  For 
intermediate barrier strengths some novel features are revealed
including a current-induced zero-bias peak and a three-humped
structure near the thermodynamical critical current density.
It is hoped that these predictions can be observed experimentally.
We conclude with the remark that this formulation can also be 
applied to the case of an d+s superconductor. Because the 
critical current for an s-wave superocnductor is larger 
than that for a d-wave one, the existence of an s component 
can be verified by a supercurrent reaching a magnitude 
between the critical values of the two waves.

We wish to thank J. Wei for helpful discussions.
This work was supported by the Texas Center for
Superconductivity and Advanced Materials at the University 
of Houston and by the Robert A. Welch Foundation (Ting).

%\end{multicols}

\end{document}